\documentclass[11pt]{article}

\usepackage{amssymb}
\usepackage{graphicx}
\usepackage{color}
\usepackage{subfigure}

\begin{document}

\baselineskip 6mm
\renewcommand{\thefootnote}{\fnsymbol{footnote}}


\newcommand{\nc}{\newcommand}
\newcommand{\rnc}{\renewcommand}



\newcommand{\tcb}{\textcolor{blue}}
\newcommand{\tcr}{\textcolor{red}}
\newcommand{\tcg}{\textcolor{green}}


\def\be{\begin{equation}}
\def\ee{\end{equation}}
\def\ba{\begin{array}}
\def\ea{\end{array}}
\def\bea{\begin{eqnarray}}
\def\eea{\end{eqnarray}}
\def\nn{\nonumber\\}


\def\ct{\cite}
\def\la{\label}
\def\eq#1{(\ref{#1})}


\def\a{\alpha}
\def\b{\beta}
\def\g{\gamma}
\def\G{\Gamma}
\def\d{\delta}
\def\D{\Delta}
\def\e{\epsilon}
\def\et{\eta}
\def\ph{\phi}
\def\Ph{\Phi}
\def\ps{\psi}
\def\Ps{\Psi}
\def\k{\kappa}
\def\l{\lambda}
\def\L{\Lambda}
\def\m{\mu}
\def\n{\nu}
\def\th{\theta}
\def\Th{\Theta}
\def\r{\rho}
\def\s{\sigma}
\def\S{\Sigma}
\def\ta{\tau}
\def\o{\omega}
\def\O{\Omega}
\def\pr{\prime}


\def\half{\frac{1}{2}}

\def\goto{\rightarrow}

\def\na{\nabla}
\def\grad{\nabla}
\def\curl{\nabla\times}
\def\div{\nabla\cdot}
\def\pa{\partial}
\def\fr{\frac}

\def\bra{\left\langle}
\def\ket{\right\rangle}
\def\lb{\left[}
\def\lc{\left\{}
\def\ls{\left(}
\def\lp{\left.}
\def\rp{\right.}
\def\rb{\right]}
\def\rc{\right\}}
\def\rs{\right)}

\def\vac#1{\mid #1 \rangle}


\def\td#1{\tilde{#1}}
\def\check{ \maltese {\bf Check!}}


\def\Tr{{\rm Tr}\,}
\def\det{{\rm det}}


\def\bc#1{\nnindent {\bf $\bullet$ #1} \\ }
\def\ch {$<Check!>$ }
\def\ss {\vspace{1.5cm}}
\def\text#1{{\rm #1}}
\def\Id{\mathds{1}}

\begin{titlepage}


\vspace{25mm}

\begin{center}
{\Large \bf Nucleon Form Factors in Nuclear Medium }

\vskip 1. cm
  {Chanyong Park$^{a,b}$\footnote{e-mail : chanyong.park@apctp.org} and Jung Hun Lee$^{a}$
\footnote{e-mail : junghun.lee@apctp.org}}

\vskip 0.5cm

{\it $^a\,$ Asia Pacific Center for Theoretical Physics, Pohang, 790-784, Korea } \\
{\it $^b\,$ Department of Physics, Postech, Pohang, 790-784, Korea }\\

\end{center}

\thispagestyle{empty}

\vskip2cm


\centerline{\bf ABSTRACT} \vskip 4mm

\vspace{1cm}

By using the AdS/CFT correspondence, we investigate various form factors between nucleons and mesons in a nuclear medium. In order to describe a nuclear medium holographically, we take into account the thermal charged AdS geometry with an appropriate IR cutoff. After introducing an anomalous dimension as a free parameter, we investigate how the nucleon's mass is affected by the change of the anomalous dimension. Moreover, we study how the form factors of nucleons rely on the properties of the nuclear medium. We show that in a nuclear medium with different numbers of proton and neutron,  the degenerated nucleon form factor in the vacuum is split to four different values depending on the isospin charges of nucleon and meson.

\vspace{2cm}


\end{titlepage}

\renewcommand{\thefootnote}{\arabic{footnote}}
\setcounter{footnote}{0}



\section{Introduction}

After the proposition of the AdS/CFT correspondence, it has been widely utilized to investigate various strongly interacting systems of quantum chromodynamics (QCD), nuclear physics, and condensed matter theory \cite{Maldacena:1997re,Witten:1998qj,Witten:1998zw,Gubser:1998bc}. In the strong coupling regime, because the traditional perturbation method does not work anymore, the AdS/CFT correspondence can shed light on understanding a variety of physical features of strongly interacting systems \cite{Aharony:1999ti,Klebanov:2000me,Horowitz:2006ct}. One of such systems we want to know is a nuclear medium in which the fundamental excitations are not quarks but their bound states called nucleon and meson \cite{Erlich:2005qh,Karch:2006pv,Sakai:2004cn,Sakai:2005yt}. In this work, we consider the holographic hard wall model on the thermal charged AdS geometry, which imitates a nuclear medium, and investigate nucleon's masses and their form factors.

In general, it is not an easy task to understand theoretically low energy physics of QCD and condensed matter physics because they are in a strong coupling regime. In a medium, moreover, the numerical method known as the lattice QCD suffers from a sign problem. In this situation, the AdS/CFT correspondence may be a useful tool to account for many interesting physical phenomena of strongly interacting systems. In the original hard wall model \cite{Erlich:2005qh}, the thermal AdS space with an IR cutoff was matched to the vacuum of the confining phase, while the AdS black hole was mapped to the deconfining phase representing the quark-gluon plasma. Various hadronic spectra and the deconfinement phase transition have been studied on this background \cite{Herzog:2006ra,DaRold:2005mxj,Hong:2007kx,Kim:2009bp}. 

In order to take further into account a nuclear medium, we must add more bulk fields which classify the properties of a nuclear medium. According to the AdS/CFT correspondence, the bulk vector field is matched to the fermionic density operator on the dual field theory side. Thus, we can mimic a nuclear medium holographically by adding additional bulk vector fields \cite{Kim:2007em}. The resulting dual gravity theory allows a new geometric solution called the thermal charged AdS (tcAdS), which is dual to the confining phase of a nuclear medium governed by hadronic spectra \cite{Lee:2009bya,Park:2009nb,Jo:2009xr}. On this background corresponding a nuclear medium, various meson's and nucleon's spectra have been studied \cite{Domokos:2007kt,Nakamura:2006xk,Park:2011zp,Lee:2013oya,Hong:2006ta,Lee:2014xda,Lee:2014gna,Lee:2015rxa,Colangelo:2010pe,Colangelo:2012jy,Mamedov:2016ype,Fadafan:2012qy,Sachan:2014jda}. These quantities are crucially associated with the two-point correlation function of nucleon and meson \cite{Park:2016hvb}. In order to understand further a nuclear medium, it is important to know the interaction between nucleons and mesons.  Those information are usually encoded into the form factor related to the three-point correlation function \cite{Grigoryan:2007vg,Hashimoto:2008zw,Brodsky:2007hb,BallonBayona:2009ar,Bayona:2010bg,Brodsky:2011xx,Nishihara:2014nsa}. In this work, we will investigate the nucleon form factors in a nuclear medium and study how the nucleon's anomalous dimension affects the nucleon's mass and form factors.


\section{Holographic description for a nuclear medium}

In general, a nuclear medium is a very complicated system including a variety of baryons and mesons together with their strong interaction. Unfortunately, there is no well-established perturbation method to understand such a strongly interacting system in traditional QFT. In this situation, the AdS/CFT correspondence may provide a new paradigm to figure out various qualitative features of strongly interacting systems in the nuclear and condensed matter theory. Even though it is not easy to construct an exact dual gravity of QCD, a variety of holographic models mimicking QCD would be useful to improve our knowledge about strongly interacting systems. In this work, we will investigate nucleon's spectra and their form factors in a nuclear medium by using the holographic hard wall model.

Let us start with summarizing briefly how we can describe a nuclear medium holographically. Proton and neutron are basic elements to represent a $U(2)$ flavor group of a nuclear medium. Since a global symmetry of the QFT maps to a local symmetry of a dual gravity according to the AdS/CFT correspondence, the dual geometry we should consider must involve a $U(2)$ local gauge symmetry.  From now on, we take into account two copies of the flavor group, $U(2)_L \times U(2)_R$, in order to clarify the parity of hadrons. Then, the eigenvalues of their Cartan subgroups provide good quantum numbers representing the quark number and isospin charge of $u$- and $d$-quark. The corresponding gravity action is given by \cite{Lee:2009bya,Park:2009nb,Jo:2009xr}
\bea
S &=&\int d^{5}x \sqrt{-G}  \left[
\frac{1}{2\kappa ^{2}}\left( \mathcal{R}-2\Lambda \right)
- \frac{1}{4g^{2}}  \ls {F}^{(L)}_{MN} {F}^{(L)MN} + {F}^{(R)}_{MN} {F}^{(R)MN} \rs \rp , \nn
&& \lp \fr{}{} \qquad \qquad \quad +\left| D_M\Phi \right|^2 +m^2\left|\Phi \right|^2 \rb
\label{2}
\eea
where $\Lambda =-6/R^{2}$ is a cosmological constant and a massive complex scalar field with $m^2 = -3/R^2$ is introduced for encoding the effect of the chiral condensate. Above the gauge field strengths and the covariant derivative are defined as
\bea
F^{(L)}_{MN} &=& \partial _{M} L_{N} -\partial _{N} L_{M} - i \lb L_M, L_N\rb , \nn
F^{(R)}_{MN} &=& \partial _{M} R_{N} -\partial _{N} R_{M} - i \lb  R_M,  R_N \rb . \nn
D_M \Ph &=& \pa_M \Ph - i L_M \Ph + i \Ph R_M . 
\eea

For describing a nuclear medium, it is sufficient to regard only the Cartan subgroups, $U(1)_L^2 \times U(1)_R^2$, because their quantum numbers classify the nuclear medium manifestly. In the holographic model, the time component of the gauge field corresponds to the density of dual particles, while the spatial component is matched to the current. Hereafter, we assume that the nuclear medium is at rest. Then, only time components of the gauge field can classify the nuclear medium we consider. Rewriting them as symmetric and antisymmetric combinations, the symmetric and antisymmetric ones represent the parity even and odd states, respectively. Since proton or neutron are identified with the lowest parity even state relying on their isospin charge, the nuclear medium composed of protons and neutrons is accomplished by taking the symmetric combination with $L_M = R_M = - V_M$ in the holographic model \cite{Lee:2013oya}. In the confining phase the fundamental excitation is not quark but nucleon, so we need to reinterpret the above quark's quantities in terms of nucleon's ones. Using the conservation of the quark number, the resulting geometry satisfying Einstein equation is given by \cite{Park:2011zp,Lee:2013oya}
\be		\la{back:geom}
ds^2 = \fr{R^2}{z^2} \ls - f(z) dt^2 + \fr{1}{f(z)} dz^2 + d \vec{x}^2  \rs ,
\ee
with
\bea
f(z) &=& 1 + \fr{3 Q^2 \k^2}{g^2 R^2} z^6 + \fr{D^2 \k^2}{3 g^2 R^2} z^6 , \nn
V^0_t &=& \fr{Q}{\sqrt{2}} \ls 2 z_{IR}^2 - 3 z^2 \rs, \nn
V^3_t &=& \fr{D}{3 \sqrt{2}} \ls 2 z_{IR}^2 - 3 z^2 \rs ,
\eea
where $Q = Q_P + Q_N$ and $D = \a Q= Q_P - Q_N$ denote the total nucleon number density and the density difference between proton and neutron. Here, $Q_p$ and $Q_N$ indicate the number density of proton and neutron respectively. This geometric solution has been known as the tcAdS geometry \cite{Lee:2009bya,Park:2009nb}. On this tcAdS background, the deconfinement phase transition and the symmetry energy have been studied \cite{Park:2011zp}. Turning on off-diagonal fluctuations of the gauge fields, they describe $SU(2)$ mesons and their spectra have been investigated in \cite{Lee:2013oya}.

When the mass of the above complex scalar field is given by $- 3/R^2$, it becomes the dual of the chiral condensate. More precisely, parameterizing the complex scalar field as
\be  \la{res:scalarmodulus}
\Ph =  \ph {\bf 1} \ e^{i \sqrt{2} \pi} ,
\ee
where ${\bf 1}$ indicates a two-by-two identity matrix, the modulus, $\ph$, is dual to the chiral condensate, while the fluctuation, $\pi$, corresponds to pion. On the above tcAdS geometry, the modulus has the following solution \cite{Lee:2013oya}
\be
\ph (z) = m_q \ z \ _2 F_1 \ls \frac{1}{6} , \half , \frac{2}{3}, - \frac{\ls D^2 + 9 Q^2 \rs
z^6 }{3 \ N_c} \rs
 + \s \ z^3 \ _2 F_1 \ls \half, \frac{5}{6},\frac{4}{3},  - \frac{\ls D^2 + 9 Q^2 \rs
z^6 }{3 \ N_c}\rs ,
\ee
where $m_q$ and $\s$ are identified with the current quark mass and chiral condensate, respectively. Here, $N_c$ denotes the rank of the gauge group. In general, the gravitational backreaction of the scalar field changes the background geometry. As shown in \cite{Lee:2010dh}, it corresponds to $1/N_c$ correction and changes the tcAdS geometry very slightly. In this note, we ignore this $1/N_c$ correction as done in the usual hard wall model.

In order to represent various mesons, we should take into accout various bulk field fluctuations, $L^a_M  \to L^a_M + l^a_M$ and $R^a_M  \to R^a_M + r^a_M$. In the axial gauge with $l^a_z=r^a_z=0$ and $l^a_t=r^a_t=0$, the left and right gauge fluctuations are further decomposed into vector and axial-vector fluctuations like \cite{Park:2011zp,Lee:2013oya,Lee:2015rxa}
\be			
l^a_m = \fr{1}{\sqrt{2}} \ls v^a_m + a^a_m \rs \quad  {\rm and } \quad
r^a_m = \fr{1}{\sqrt{2}} \ls v^a_m - a^a_m \rs ,
\ee
where $m$ indicates one of the spatial directions. In order to identify them with the vector mesons, we need to redefine the vector fluctuations as the form showing their flavor charge explicitly
\be		\la{def:vector}
v^a_m = \r^0_m   \quad  , \quad  v^1_m =  \fr{1}{\sqrt{2}}  \ls \r^{+}_m + \r^{-}_m  \rs
\quad  {\rm and } \quad  v^2_m =  \fr{i}{\sqrt{2}}  \ls \r^{+}_m - \r^{-}_m  \rs .
\ee
Then, $\r^i_m$ indicates $\r$-meson with the $i$ flavor charge. Similarly, axial-vector and pion are also rewritten as the form representing the SU(2) charge manifestly
\bea        \la{def:avector}
a^a_m = a^0_m   \quad  , \quad  a^1_m &=&  \fr{1}{\sqrt{2}}  \ls a^{+}_m + a^{-}_m  \rs
\quad  , \quad  a^2_m =  \fr{i}{\sqrt{2}}  \ls a^{+}_m - a^{-}_m  \rs , \nn
\pi^3 = \pi^0   \quad  , \quad  \pi^1 &=&  \fr{1}{\sqrt{2}}  \ls \pi^{+} + \pi^{-} \rs  \quad , 
\quad  {\rm and } \quad  \pi^2 =  \fr{i}{\sqrt{2}}  \ls \pi^{+} - \pi^{-}  \rs .
\eea
Above the axial-vector field usually couples to the pseudoscalar field. Using the following gauge transformation
\be        \la{def:agaugefix}
a^a_m = \bar{a}^a_m + \pa_m \chi^a ,
\ee
together with imposing $0 = \pa^m \bar{a}^a_m$, they are decoupled and the longitudinal mode of the axial-vector represents another pseudoscalar field, $\chi$. Their spectra and form factors in the nuclear medium have already been studied in \cite{Lee:2014gna,Park:2016hvb}.

For later use, we summarize the equations of motion for mesons \cite{Lee:2014gna,Park:2016hvb}. After the Fourier mode expansion, the mode function of $\r$-meson should satisfy
\bea			\la{eq:vectormeson}
0 &=&   \pa_z^2 f_{\r}^{(0)}
+ \frac{ -3 N_c + 5 (D^2 + 9 Q^2) z^6 }{  z \lb 3 N_c + (D^2 + 9 Q^2) z^6 \rb} \pa_z f_{\r}^{(0)}
+ \frac{9 N_c^2 \ls \o_\r^{(0)}\rs^2 }{ \lb 3 N_c + (D^2 + 9 Q^2) z^6) \rb^2 }  f_{\r}^{(0)} , \nn
0 &=&   \pa_z^2 f_{\r}^{(\pm)} 
+ \frac{ -3 N_c + 5 (D^2 + 9 Q^2) z^6 }{ z \lb 3 N_c + (D^2 + 9 Q^2) z^6 \rb } \pa_z 
f_{\r}^{(\pm)}
+ \frac{9 N_c^2 \ls \o_\r^{(\pm)} \mp V_0^3 \rs^2 }{ \lb 3 N_c + (D^2 + 9 Q^2) z^6) \rb^2 }  
f_{\r}^{(\pm)}  ,
\eea
where the superscript, $(i)$, of the mode function, $f_{\r}^{(i)}$, indicates the isospinc charge of $\r$-meson. Similarly, the mode function of $a_1$ axial-vector meson satisfies
\bea
0 &=& \pa_z^2 f_a^{(0)}
+ \frac{ 5 (D^2 + 9 Q^2) z^6 -3 N_c  }{  z \lb 3 N_c + (D^2 + 9 Q^2) z^6) \rb}  \pa_z f_a^{(0)} \nn
&&  + \frac{9 N_c^2 \ls \o_a^{(0)}\rs^2 z^2 -  12 N_c \ g^2  \lb 3 N_c + (D^2 + 9 Q^2) z^6 \rb \ph^2 }{z^2 \lb 3 N_c + (D^2 + 9 Q^2) z^6) \rb^2 }   f_a^{(0)} , \nn
0 &=& \pa_z^2 f_a^{(\pm)}
+ \frac{ 5 (D^2 + 9 Q^2) z^6 -3 N_c  }{   z \lb 3 N_c + (D^2 + 9 Q^2) z^6) \rb} \pa_z f_a^{(\pm)}   \nn
&& + \frac{9 N_c^2 (\o_a^{(\pm)}  \mp V_0^3 )^2 z^2   -  12 N_c  g^2 \lb 3 N_c + (D^2 + 9 Q^2) z^6 \rb \ph^2 }{z^2 \lb 3 N_c + (D^2 + 9 Q^2) z^6) \rb^2 }   f_a^{(\pm)} , 
\eea
In the axial gauge mentioned before, pion's mode function is governed by the following equations 
\bea
\frac{z^3 f(z)}{g^2\phi^2} \partial_z\left(\frac{g^2\phi^2 f(z)}{z^3}\partial_z f_{\pi}^{(0)}\right) 
&=&  \left( \ls \o_\pi^{(0)} \rs^2-f(z) |\vec{p}|^2\right)\left( f_{\chi}^{(0)} -  f_{\pi}^{(0)}\right)  ,\nonumber\\
 \frac{z^3 f(z)} {4g^2\phi^2}\partial_z\left(\frac{ \ls \o_\pi^{(0)} \rs^2-f(z) |\vec{p}|^2}{z}\partial_z f_{\chi}^{(0)}\right)
&=&\left( \ls \o_\pi^{(0)} \rs^2- f(z)  |\vec{p}|^2\right)\left(f_{\chi}^{(0)} -  f_{\pi}^{(0)}\right)   ,\nonumber\\
 \frac{z^3 f(z)} {g^2\phi^2}\partial_z\left(\frac{g^2\phi^2 f(z)}{z^3}\partial_z f_{\pi}^{(\pm)}\right)  
 &=& \left( \ls \o_\pi^{(\pm)} \rs^2 \mp {V}^3_t \o_\pi^{(\pm)}  - f(z) |\vec{p}|^2\right) f_{\chi}^{(\pm)} \nn
 && - \left( \left( \o_\pi^{(\pm)} \mp{V}^3_t\right)^2 -f(z) |\vec{p}|^2\right) f_{\pi}^{(\pm)}  ,\nonumber\\
\frac{z^3 f(z)} {4g^2\phi^2}\partial_z\left(\frac{ \ls \o_\pi^{(\pm)} \rs^2-f(z) |\vec{p}|^2}{z}\partial_z f_{\chi}^{(\pm)}\right)  
&=& \left( \ls \o_\pi^{(\pm)} \rs^2-f(z) |\vec{p}|^2 +\frac{ \ls {V}^3_t\rs^2 z^2}{4g^2\phi^2}|\vec{p}|^2\right) f_{\chi}^{(\pm)} \nn
&& -\left( \ls \o_\pi^{(\pm)} \rs^2 \mp{V}^3_t   \o_\pi^{(\pm)}  - f(z)  |\vec{p}|^2\right) f_{\pi}^{(\pm)} ,
\eea
where $\o_{q}^{(i)}$ and $\vec{p}$ indicate the energy and momentum of $q$-meson with an isospin charge $(i)$. Above $\pi$ and $\chi$ indicate fluctuations of the scalar field and longitudinal component of the axial-vector, respectively. Solving above equations with the proper boundary conditions, Dirichlet boundary condition at the asymptotic boundary and Neumann boundary condition at the IR cutoff denoted by $z_{IR}$, determines the mass of various mesons \cite{Lee:2014xda,Lee:2014gna}.

Note that the above mode functions should be appropriately normalized to define form factors. From the dual gravity, $\r$ and $a_1$-mesons have the following normalization  \cite{Park:2016hvb}
\be
1 = \int_0^{z_{IR}} dz \ \fr{\ls f_\r^{(a)} \rs^2}{z f(z)}  \quad {\rm and} \quad 1 = \int_0^{z_{IR}} dz \ \fr{\ls f_a^{(a)} \rs^2}{z f(z)}
\ee
On the other hand, $\pi$ and $\chi$ representing pion have different normalizations due to their different origins on the dual gravity side. Concerning their gravity origin, $\pi$ and $\chi$ have the following normalization
\be
1 = \int_0^{z_{IR}} dz \ \fr{\ls f_\chi^{(a)} \rs^2}{z f(z)} \quad {\rm and} \quad 1 = \int_0^{z_{IR}} dz \ \fr{\ls f_\pi^{(a)} \rs^2}{z^3 f(z)} .
\ee
When we evaluate form factors of nucleons in the next section, we will utilize these normalizations.

\section{Nucleons in the nuclear medium}

In order to imitate nucleons holographically, one should take into account Dirac fermions on the tcAdS space, which are governed by \cite{Hong:2006ta,Lee:2014xda,Lee:2015rxa}
\begin{eqnarray}		\la{act:fermion}
S &=& i \int d^{5}x\sqrt{- G}\left[  \overline{\Psi }^{1} \Gamma ^{M}\nabla_{M}\Psi^{1}
+ \overline{\Psi }^{2} \Gamma ^{M}\nabla_{M}\Psi^{2}
- m_{1}\overline{\Psi}^{1}\Psi^{1}-m_{2}\overline{\Psi }^{2}\Psi^{2} \rp \nn
&& \qquad \qquad \qquad - \lp g_Y  \ls
\overline{\Psi}^{1} \Ph \Psi^{2} + \overline{\Psi}^{2} \Ph^{+} \Psi^{1} \rs
\right] . \label{42}
\end{eqnarray}
Here, $\Psi^{1}$ and $\Psi^{2}$ transform as $\ls \frac{1}{2},0 \rs$ and $\ls 0,\frac{1}{2} \rs$ under the flavor group $U(2)_{L}\times U(2)_{R}$ and the Yukawa term breaks the chiral symmetry. In the AdS/CFT context, the conformal dimension of nucleon is related to the mass of the bulk fermion  \cite{Hong:2006ta,Henningson:1998cd,Muck:1998rr,Contino:2004vy}
\be
m^2 = \ls \D - 2 \rs^2 ,
\ee
If there is no anomalous dimension, the nucleon's conformal dimension is given by $9/2$ at a conformal fixed point. In this case, the mass of the bulk fermion should be $\pm 5/2$. However, the existence of an anomalous dimension significantly modifies the nucleon's conformal dimension. Although it would be interesting to clarify what the anomalous dimension is, it goes beyond the scope of this work. We leave this issue as a future work. In this work, we treat the nucleon's conformal dimension as a free parameter due to the undetermined anomalous dimension. In order to realize the chirality, furthermore, we focus on the case with $m_1 = - m_2= \D-2 $ for $\D > 2$.

Following the conventions in \cite{Lee:2015rxa}, the Dirac equation is reduced to
\begin{eqnarray}		\la{eq:equationferm}
0 &=& \lb e^M_C \G^C \left(\partial _{M} - \frac{i}{4}\omega _{M}^{AB}\Gamma _{AB} +
i  V_M \right) - m_{1} \rb \Psi^{1}
- g_Y \ph \Ps^{2} , \nn
0 &=& \lb e^M_C \G^C \left(\partial _{M} - \frac{i}{4}\omega _{M}^{AB}\Gamma _{AB} +
i  V_M \right) - m_{2} \rb \Psi^{2}
- g_Y \ph \Ps^{1} .
\end{eqnarray}%
Under the following Fourier mode expansion 
\be
\Ps^{i} (z,t,\vec{x}) = \int \fr{ d^4 p}{(2 \pi)^4} \ \Ps^i (z,\o,\vec{p}) \ e^{- i  \ls \o t - \vec{p} \cdot \vec{x} \rs }  \quad \ls i=1,2 \rs ,
\ee
the fermionic mode function can be classified as  
\be			\la{def:bulkfermion}
\Ps^{1} (z,\o,\vec{p}) = \ls \begin{array}{c}
f_L^{1(n,\pm,\pm)} \ \ps^{(n,\pm,\pm)}_L\\
f_R^{1(n,\pm,\pm)}  \ \ps^{(n,\pm,\pm)}_R
\end{array} \rs  \quad {\rm and} \quad
\Ps^{2} (z,\o,\vec{p})  = \ls \begin{array}{c}
f_L^{2(n,\pm,\pm)} \ \ps^{(n,\pm,\pm)}_L\\
f_R^{2(n,\pm,\pm)}  \ \ps^{(n,\pm,\pm)}_R
\end{array} \rs ,
\ee
where $\ps_L$ and $\ps_R$ mean $4$-dimensional Weyl spinors satisfying $\ps_L = \g^5 \ps_L$ and $\ps_R = - \g^5 \ps_R$. Above $(n,\pm,\pm)$ denotes the $n$-th resonance, parity, and isospin charge respectively. In this case, the combination satisfying the following condition \cite{Kim:2009bp,Hong:2006ta,Lee:2014xda,Lee:2015rxa}
\be		\la{con:symmetric}
 f_L^{1(n,+,\pm)} = f_R^{2(n,+,\pm)} \ \text{and} \quad  f_R^{1(n,+,\pm)} = - f_L^{2(n,+,\pm)} ,
\ee
describes a parity even state, while a parity odd state is represented as 
\be   \la{con:antisymmetric}
f_L^{1(n,-,\pm)} = - f_R^{2(n,-,\pm)} \ \text{and} \quad  f_R^{1(n,-,\pm)} = f_L^{2(n,-,\pm)} .
\ee
Since proton and neutron we are interest in appear in the lowest parity even state, from now on we concentrate on the case with $(1,+,\pm)$ where the last sign indicates the isospin charge of proton and neutron.

To solve the Dirac equations, firstly we need to determine mode functions as eigenfunctions of $\vec{\s}  \cdot \vec{p}$. The matrix $\vec{\s}  \cdot \vec{p}$ generally allows two eigenvalues, $\pm p$, with $p = \left| \vec{p} \right|$. In this case, the mode function can be identified with one of two eigenfunctions up to
normalization. For later consistency, let us focus on the case in which $f^1_L$ and $f^1_R$ are
eingenfuntions with the eigenvalue $p$ while the eigenvalues of $f^2_L$ and $f^2_R$ are given
by $-p$.  Then, the bulk Dirac equation for the parity even state reduces to
\bea		\la{eq:protonneutroneq}
\ls \begin{array}{cc}
{\cal D}_-    & \fr{g_Y \ph}{z}   \\
\fr{g_Y \ph}{z}   &  {\cal D}_+  
\end{array} \rs
\ls \begin{array}{c}
f_L^{1(n,+,\pm)} \\
f_R^{1(n,+,\pm)}
\end{array} \rs
&=&
\ls \begin{array}{cc}
- E_+  & 0 \\
0 & E_-
\end{array} \rs
\ls \begin{array}{c}
f_R^{1(n,+,\pm)} \\
f_L^{1(n,+,\pm)}
\end{array} \rs .
\eea
with
\bea
{\cal D}_{\pm} &=&  \sqrt{f(z)} \lb \pa_z - \fr{2}{z} \ls 1 - \fr{z f'}{8  f(z) } \rs \rb \pm \fr{\D-2}{z} , \\
E_{\pm} &=&  \fr{1}{\sqrt{f(z)}}  \ls \o - V_t \rs  \pm  p  \ 
\la{eq:matrixeq} .
\eea
Depending on the isospin charge, this equation further splits into two cases. Proton with the positive isospin charge  is governed by
\bea		\la{eq:protoneq}
&& \ls \begin{array}{cc}
 {\cal D}_-  & \fr{g_Y \ph}{z}   \\
\fr{g_Y \ph}{z}   &  {\cal D}_+  
\end{array} \rs
\ls \begin{array}{c}
f_L^{1(1,+,+)} \\
f_R^{1(1+,+)}
\end{array} \rs  \nn
&& \qquad \qquad=
\ls \begin{array}{cc}
-  \lc \fr{1}{\sqrt{f(z)}}  \ls \o - \fr{V^0_t  + V_t^3}{2} \rs  +  p   \rc   & 0 \\
0 & \fr{1}{\sqrt{f(z)}}  \ls \o - \fr{V^0_t  + V_t^3}{2} \rs  - p
\end{array} \rs
\ls \begin{array}{c}
f_R^{1(1,+,+)} \\
f_L^{1(1,+,+)}
\end{array} \rs ,
\eea
while \eq{eq:protonneutroneq} for neutron  with the negative isospin charge becomes
\bea		\la{eq:neutroneq}
&& \ls \begin{array}{cc}
 {\cal D}_-    &  \fr{g_Y \ph}{z}   \\
 \fr{g_Y \ph}{z}   &  {\cal D}_+  
\end{array} \rs
\ls \begin{array}{c}
f_L^{1(1,+,-)} \\
f_R^{1(1,+,-)}
\end{array} \rs  \nn
&& \qquad \qquad=
\ls \begin{array}{cc}
-  \lc \fr{1}{\sqrt{f(z)}}  \ls \o - \fr{V^0_t  - V_t^3}{2} \rs  +  p   \rc   & 0 \\
0 & \fr{1}{\sqrt{f(z)}}  \ls \o - \fr{V^0_t - V_t^3}{2} \rs  - p
\end{array} \rs
\ls \begin{array}{c}
f_R^{1(1,+,-)} \\
f_L^{1(1,+,-)}
\end{array} \rs .
\eea

Taking $V^0_t =0$, $V^3_t = \text{const}$ and $f(z)=1$, the above equations reduce to those for nucleons in the isospin medium \cite{Lee:2014xda}. In the nuclear medium, unlike the isospin medium, the energy and mass of nucleons crucially depends on the nuclear medium due to the nontrivial radial coordinate dependence of the metric and background gauge fields. For $\D=9/2$, nucleon's spectra and their dispersion relations have been studied in \cite{Lee:2015rxa} where the Yukawa coupling was taken by $g_Y = 4.699$ in order to fix nucleon's mass to be $940$MeV in the vacuum with $Q=0$ and $p=0$. Similarly, when we consider the anomalous dimension we should take a different Yukawa coupling to obtain the known nucleon's mass in the vacuum. In Table 1, we show several corresponding Yukawa couplings. In Fig. 1, we plot the nucleon mass relying on the conformal dimension of the nucleon. The result shows that the nucleon mass increases as its conformal dimension decreases.

 \begin{table}
\begin{center}
\begin{tabular}{|c||c|c|c|c|c|}
\hline
$\D$ & 9/2 & 4 & 7/2 & 3 & 5/2 \\
\hline 
$g_Y$ & 4.699 & 5.029 & 5.619  & 7.081 & 10.727 \\
\hline
\end{tabular}
\end{center}
\vspace{-0.3cm}\caption{Yukawa couplings depending on the anomalous dimension which fix the nucleon's mass to be $940$MeV in the vacuum. } 
\label{default}
\end{table}%

\section{Nucleon form factors in the nuclear medium}

Now, let's investigate nucleon form factors in the nuclear medium. In general, since nucleon and meson have infinitely many resonances,  infinitely many nucleon form factors are possible. In this work, we concentrate only on the lowest resonances because they are dominant in the low energy limit. Proton and neutron we are interested in are the lowest nucleons. From the previous fermionic action in \eq{42}, the interaction terms at cubic order are given by
\bea
S_{int} &=&  \int d^{5}x\sqrt{-G}\left[  \overline{\Psi }^{1}  \Gamma ^{M}
l_M \Psi^{1}
+ \overline{\Psi }^{2}  \Gamma^{M} r_{M} \Psi^{2}
+ \sqrt{2} g_Y \ph \ls
\overline{\Psi}^{1} \pi \Psi^{2} - \overline{\Psi}^{2} \pi \Psi^{1} \rs
\right] ,
\end{eqnarray}
where $l_M$ and $r_M$ are fluctuations of the left and right flavor group. Rewriting it in terms of the vector and axial-vector fluctuations gives rise to
\bea
S_{int} &=& \int d^5 x \lb \fr{1}{\sqrt{2} z^4} \ls \overline{\Psi }^{1} \gamma ^{m}
v_m \Psi^{1} + \overline{\Psi }^{2} \gamma ^{m} v_m \Psi^{2}  \rs
+ \fr{1}{\sqrt{2} z^4} \ls \overline{\Psi }^{1} \gamma ^{m}
\bar{a}_m \Psi^{1} - \overline{\Psi }^{2} \gamma ^{m} \bar{a}_m \Psi^{2}  \rs  \rp \nn
&& \qquad  \qquad \lp + \fr{1}{\sqrt{2} z^4} \ls \overline{\Psi }^{1} \gamma ^{m}
\pa_m  \chi \Psi^{1} - \overline{\Psi }^{2} \gamma ^{m} \pa_m  \chi \Psi^{2}  \rs
+ \fr{\sqrt{2} g_Y \ph}{z^5} \ls
\overline{\Psi}^{1} \pi \Psi^{2} - \overline{\Psi}^{2} \pi \Psi^{1} \rs
\rb  ,
\eea
where $v_m = v_m^i \s^i$, $\bar{a}_m= \bar{a}_m^i \s^i$, $\chi= \chi^i \s^i$, and $\pi= \pi^i \s^i$ with the SU(2) flavor group generator, $\s^i$.
The first line expresses the interactions of nucleons with vector and axial-vector mesons,
while the last line shows the interaction with pseudoscalar mesons.
Since the lowest nucleons are in the parity even state as mentioned before, the bulk Dirac fermion in \eq{def:bulkfermion} can be recombined into the  Dirac fermion of the dual field theory
\be
\ps^{(1,+,\pm)}  = \ls \begin{array}{c}
 \ps^{(1,+,\pm)}_L\\
 \ps^{(1,+,\pm)}_R
\end{array} \rs .
\ee
Depending on its SU(2) charge, it becomes either proton, $P = \ps^{(1,+,+)}$, or
neutron, $N = \ps^{(1,+,-)}$ where the SU(2) charge, $\pm$, implies $\pm 1/2$.
For convenience, we rewrite the mode functions only in terms of the SU(2) charge like $f_L^{(\pm 1/2)}$ and $f_{meson}^{{(charge)}}$ where $(charge)$ means $0$ and $\pm 1$ for mesons. In addition, the mode functions of nucleons are normalized as \cite{Lee:2014xda}
\bea
1 = \int_0^{z_{IR}} \fr{dz}{z^4 \sqrt{f}}  \ls \left| f_L^{(i)} \right|^2   + \left| f_R^{(i)}  \right|^2\rs  .
\eea

\begin{figure}
\begin{center}
\vspace{-0cm}
\hspace{-0cm}
\subfigure[]
{ \includegraphics[angle=0,width=0.48\textwidth]{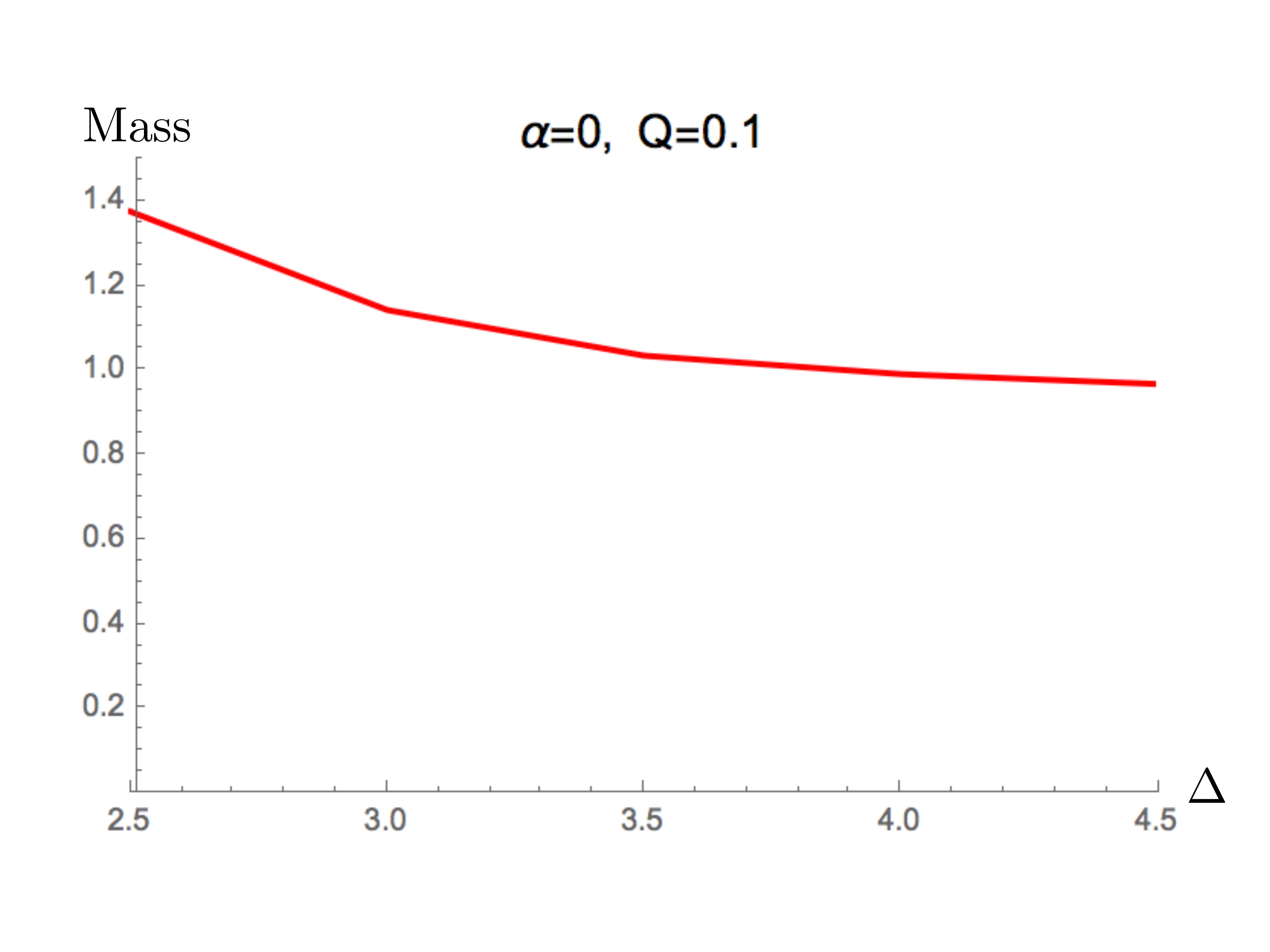}}
\hspace{0cm}
\subfigure[]
{ \includegraphics[angle=0,width=0.48\textwidth]{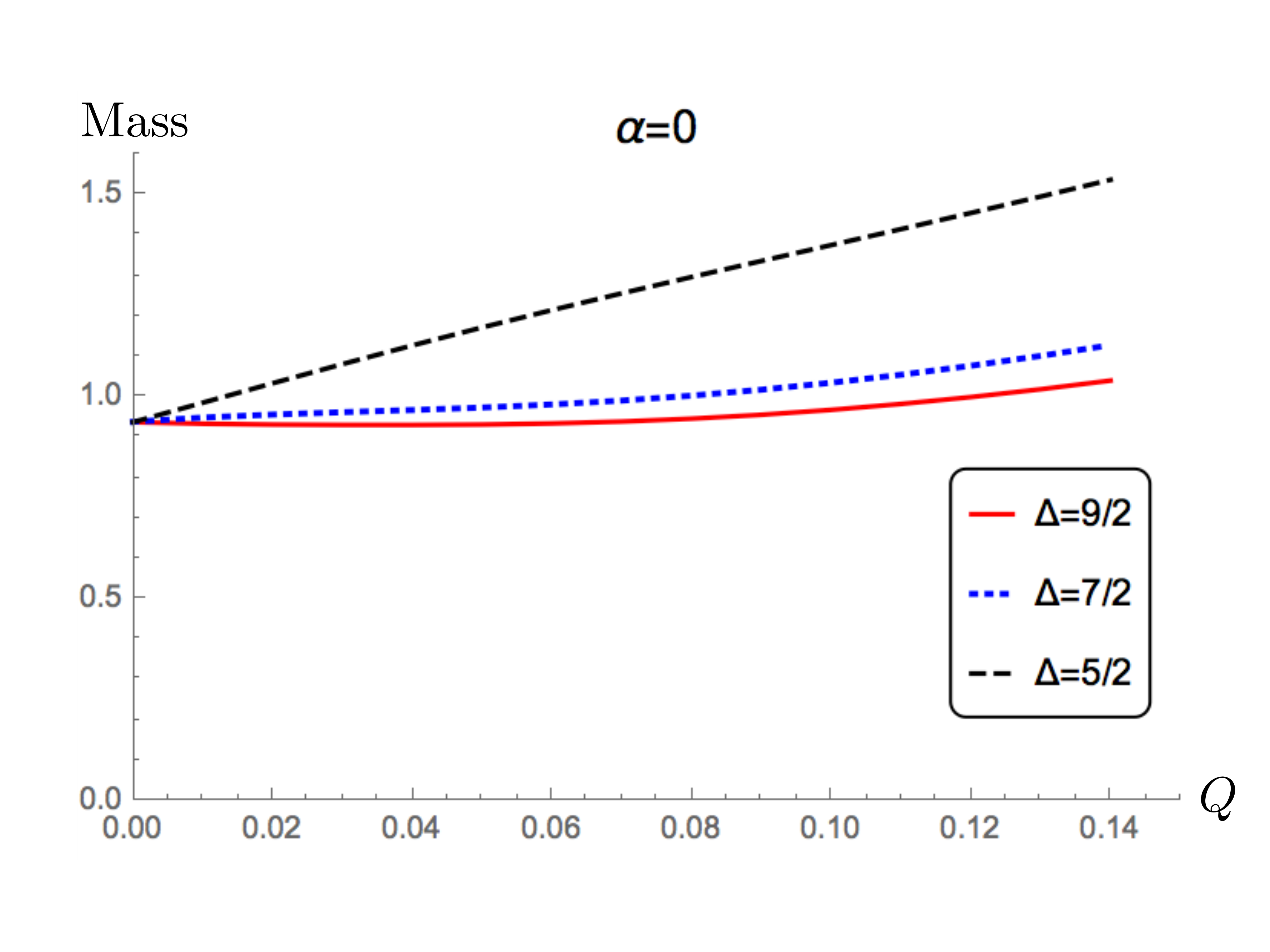}}
\vspace{-0cm}
\caption{\small  Nucleon's mass depending on $\D$ and $Q$. In (a) where we take $Q=0.1$, the nucleon mass decreases as $\D$ increases. The nucleon mass with different anomalous dimension is depicted in (b). }
\label{number}
\end{center}
\end{figure}

Then, the interaction between nucleon and $\r$-meson is represented as
\bea
S_{\r} &=& \int d^4 x \ \lb  \ \fr{1}{\sqrt{2} } \ g_{\r} \ls - 1/2 ,0,1/2\rs  \  \overline{P} \g^m \r^0_m P
-   \fr{1}{\sqrt{2} } \ \ g_{\r} \ls 1/2 ,0,-1/2\rs  \  \overline{N} \g^m \r^0_m N \rp  \nn
&& \qquad  \qquad \lp +  \ g_{\r} \ls - 1/2 ,+1,-1/2\rs  \  \overline{P} \g^m \r^+_m N
+   \ g_{\r} \ls 1/2 ,-1,1/2\rs  \  \overline{N} \g^m \r^-_m P  \ \rb  ,
\eea
where the form factor is defined as
\be
g_{\r} \ls - i,j,k\rs = \int dz \  \fr{1}{2 z^4} \
\ls f_L^{(i) *} f_{\r}^{(j)} f_L^{(k)} + f_R^{(i) *} f_{\r}^{(j)} f_R^{(k)}  \rs .
\ee
Note that due the conservation law of the SU(2) flavor charge,  sum of arguments in the nucleon form factor must vanish.
For the $a_1$-meson, similarly, the form factor is given by
\be
g_{a} \ls - i,j,k\rs = \int dz \  \fr{1}{2 z^4} \
\ls f_L^{(i) *} f_{a}^{(j)} f_L^{(k)} - f_R^{(i) *} f_{a}^{(j)} f_R^{(k)}  \rs ,
\ee
and the interaction with nucleons is governed by
\bea
S_{a} &=& \int d^4 x \ \lb \fr{1}{\sqrt{2}}  \ g_{a} \ls - 1/2 ,0,1/2\rs  \  \overline{P}
\g^m  \g^5 a^0_m P
- \fr{1}{\sqrt{2}}  \ g_{a} \ls 1/2 ,0,-1/2\rs  \  \overline{N} \g^m  \g^5 a^0_m N \rp  \nn
&& \qquad  \qquad \lp + \ g_{a} \ls - 1/2 ,+1,-1/2\rs  \  \overline{P} \g^m  \g^5 a^+_m N
+ \ g_{a} \ls 1/2 ,-1,1/2\rs  \  \overline{N} \g^m  \g^5 a^-_m P  \ \rb  .
\eea
Finally, the interactions with pseudoscalar mesons are described by
\bea
S_{sc} &=& \int d^4 x \ \lb \fr{1}{\sqrt{2}}  \ g_{\chi} \ls - 1/2 ,0,1/2\rs  \  \overline{P} \g^m  \g^5 \pa_m \chi^0 P
- \fr{1}{\sqrt{2}}   \ g_{\chi} \ls 1/2 ,0,-1/2\rs  \  \overline{N} \g^m \g^5 \pa_m \chi^0 N \rp  \nn
&& \qquad  \qquad  + 2  \ g_{\chi} \ls - 1/2 ,+1,-1/2\rs  \  \overline{P} \g^m  \g^5
\pa_m \chi^+  N
+ 2  \ g_{\chi} \ls 1/2 ,-1,1/2\rs  \  \overline{N} \g^m  \g^5 \pa_m \chi^-  P  \  \nn
&& \qquad  \qquad - \fr{1}{\sqrt{2}}   \ g_{\pi} \ls - 1/2 ,0,1/2\rs  \  \overline{P} \g^5 \pi^0 P
+ \fr{1}{\sqrt{2}}   \ g_{\pi} \ls 1/2 ,0,-1/2\rs  \  \overline{N}  \g^5  \pi^0  N   \nn
&& \qquad  \qquad \lp -  \ g_{\pi} \ls - 1/2 ,+1,-1/2\rs  \  \overline{P}  \g^5 \pi^+  N
-  \ g_{\pi} \ls 1/2 ,-1,1/2\rs  \  \overline{N} \g^5    \pi^-  P  \ \rb  ,
\eea
where the form factor between nucleons and pion is determined by
\bea
g_{\chi} \ls - i,j,k\rs &=& \int dz \ \fr{1}{2 z^4} \
\ls f_L^{(i) *} f_{\chi}^{(j)} f_L^{(k)} - f_R^{(i) *} f_{\chi}^{(j)} f_R^{(k)}  \rs  , \nn
g_{\pi} \ls - i,j,k\rs &=& \int dz \ \fr{ g_Y \ph}{z^5} \
\ls f_L^{(i) *} f_{\pi}^{(j)} f_L^{(k)} + f_R^{(i) *} f_{\pi}^{(j)} f_R^{(k)}  \rs  .
\eea

\begin{figure}[h!]
\begin{center}
\vspace{-0cm}
\hspace{-0cm}
\subfigure[]
{ \includegraphics[angle=0,width=0.48\textwidth]{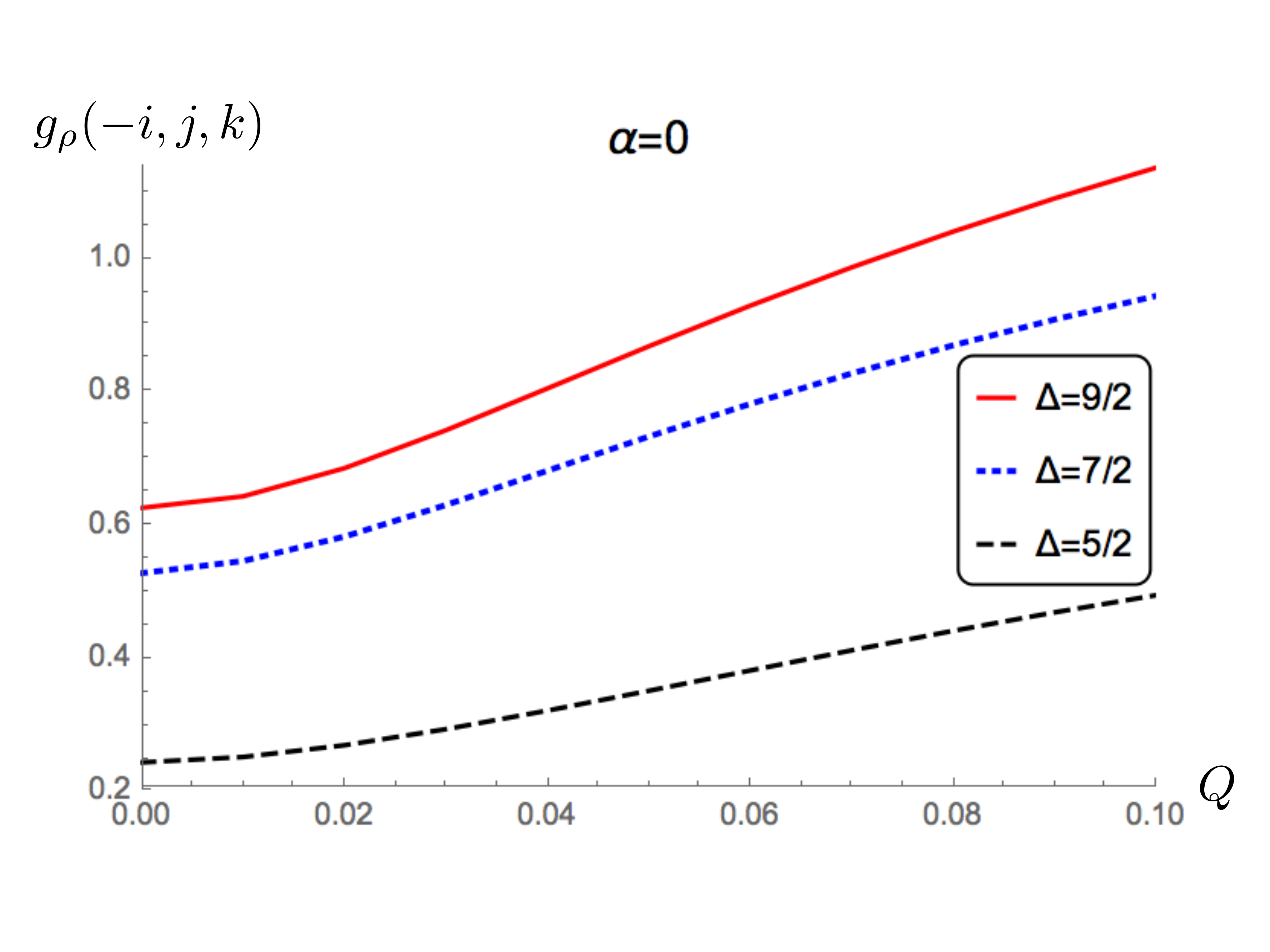}}
\hspace{0cm}
\subfigure[]
{ \includegraphics[angle=0,width=0.48\textwidth]{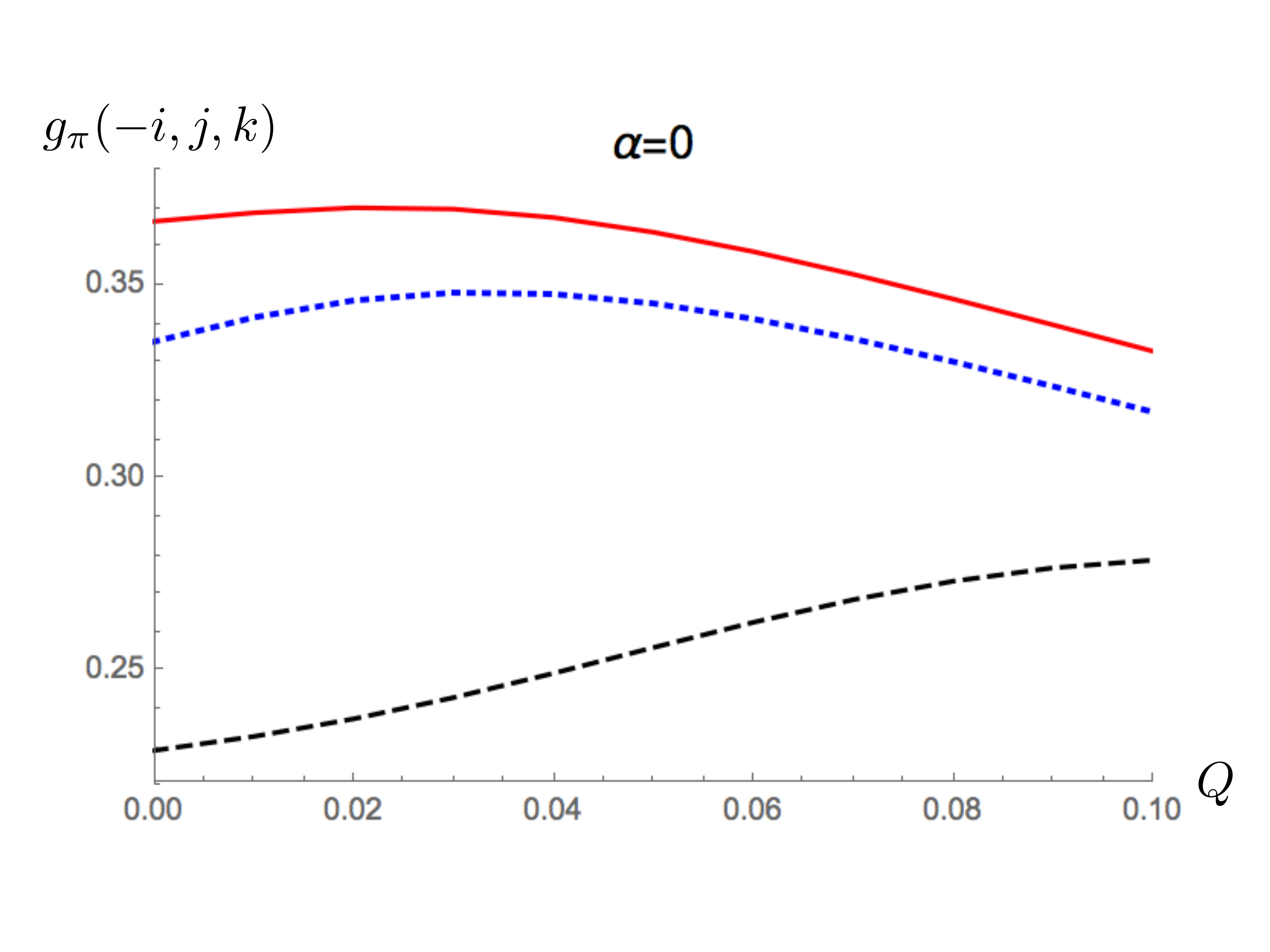}}
\vspace{-0cm}
\subfigure[]
{ \includegraphics[angle=0,width=0.48\textwidth]{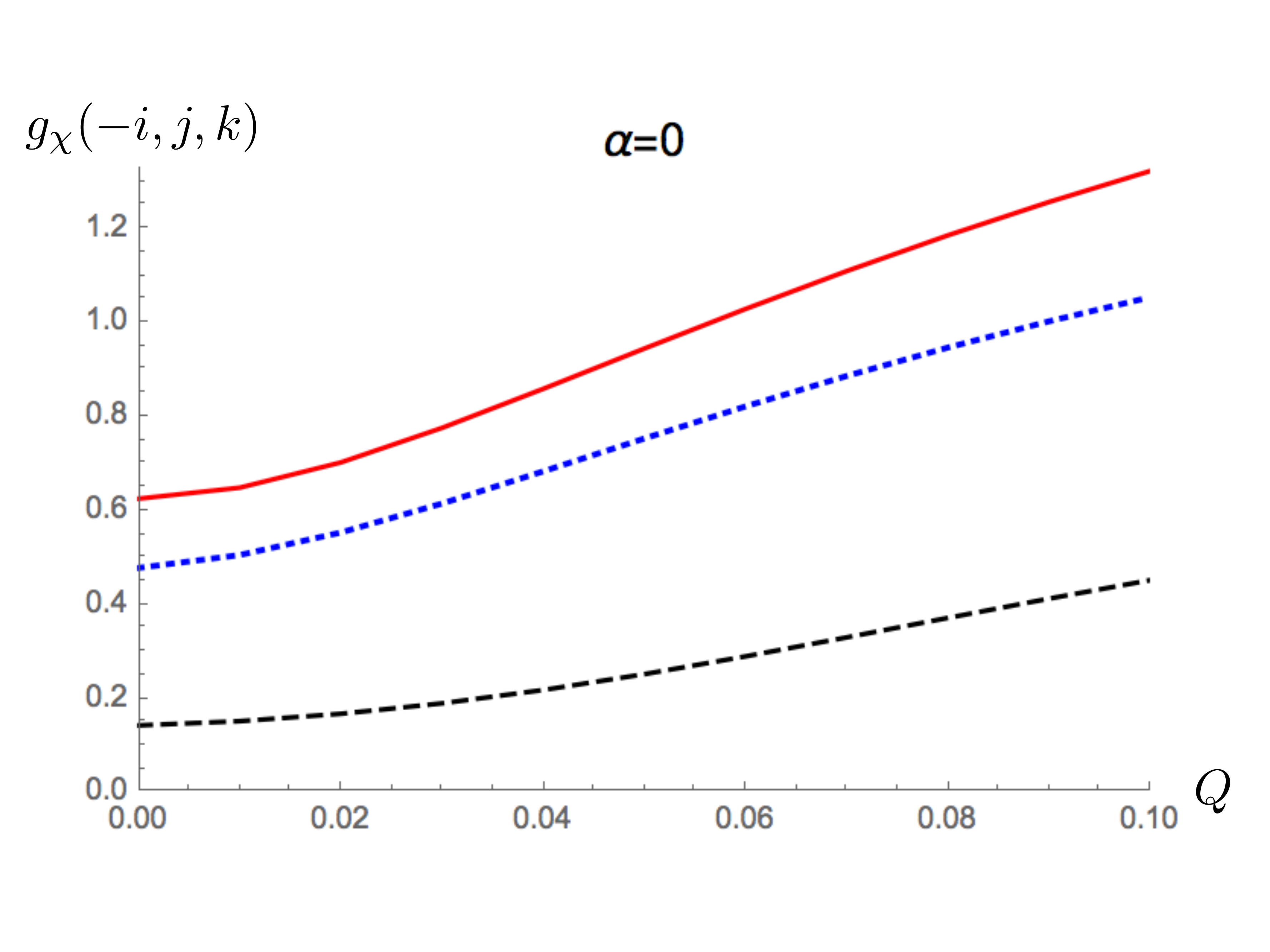}}
\vspace{-0cm}
\caption{\small  Degenerated form factors depending on $\D$ and $Q$ with $\alpha=0$. In the figure (a) and (c), the nucleon form factors, $g_\rho$ and $g_\chi$, increase as $Q$ increases. In the figure (b), one can see that there are two different behaviors : For $\Delta=9/2$, $7/2$, the form factors $g_\pi$ increases slightly and decreases with increasing $Q$. On the other hand, the form factor with $\Delta=5/2$ increases as $Q$ increases even in the high density regime.}
\label{number}
\end{center}
\end{figure}

When the isospin interaction is turned off, we depict the form factors between nucleons and mesons in Fig. 2. Because of the absence of the isospin interaction, the form factors of charged and neutral particles become degenerate. In Fig. 2, the result shows that the effect of the anomalous dimension denoted by $9/2-\D$ usually reduces the strength of the form factor except $g_\pi$ in the high density region. The numerical result also shows that the form factor between nucleon and $\r$-meson increases as the density of the nuclear medium increases. On the other hand, the nucleon's form factor with pseudoscalar meson shows two different behaviors for a small anomalous dimension. When the density of the nuclear medium increases, the nucleon's form factor with $\pi^{i}$ decreases, while that with $\chi^i$ oppositely increases. This feature is crucially modified for a large anomalous dimension with a small conformal dimension. In the case with $\D=5/2$, the form factor with $\pi^{i}$ also increases as the density of the nuclear medium increases. This fact indicates that the anomalous dimension can affect a significant effect on the behavior of the form factor with the pseudoscalar meson. Therefore, it would be interesting to evaluate the anomalous dimension of nucleon correctly by studying the holographic renormalization group flow. We leave this issue as a future work.

If we further consider the isospin interaction, the degeneracy of the form factors mentioned before splits because the isospin interaction distinguishes the isospin charge of nucleons and mesons. In Fig. 3, we consider the case with $\a=-1/2$ which implies that the nuclear medium we considered is composed of $75\%$ neutrons and $25\%$ protons. The results in Fig. 3 show that the degenerated form factor for $\a=0$ splits four different cases relying on the isospin charges of nucleons and mesons. If the charges of nucleons are given, the charge of the remaining meson is automatically determined due to the charge conservation. This is the reason why we obtain four non-degenerate form factors when turning on the isospin interaction. More precisely, let us take into account the form factors having the same field contents, $ \overline{P}  \g^5 \pi^+  N$ and $\overline{N} \g^5    \pi^-  P$,  which are the complex conjugates to each other. Even in this case, the result in Fig. 3 shows that the corresponding form factors, $g_{\pi} \ls 1/2 ,-1,1/2\rs$ and $g_{\pi} \ls - 1/2 ,+1,-1/2\rs$, have different values. This is because the background nuclear medium we considered has different number densities of proton and neutron. This asymmetry of the nucleon's number density in the nuclear medium causes the split of the form factor and we finally obtain four different form factors as mentioned before. In general, our numerical results show that the nucleon form factor in the nuclear medium crucially relies on the properties of the nuclear medium as expected. So it would be interesting and important to compare our holographic result with data obtained from the future particle experiments.

\begin{figure}
\begin{center}
\vspace{-1cm}
\hspace{-0cm}
\subfigure[]
{ \includegraphics[angle=0,width=0.48\textwidth]{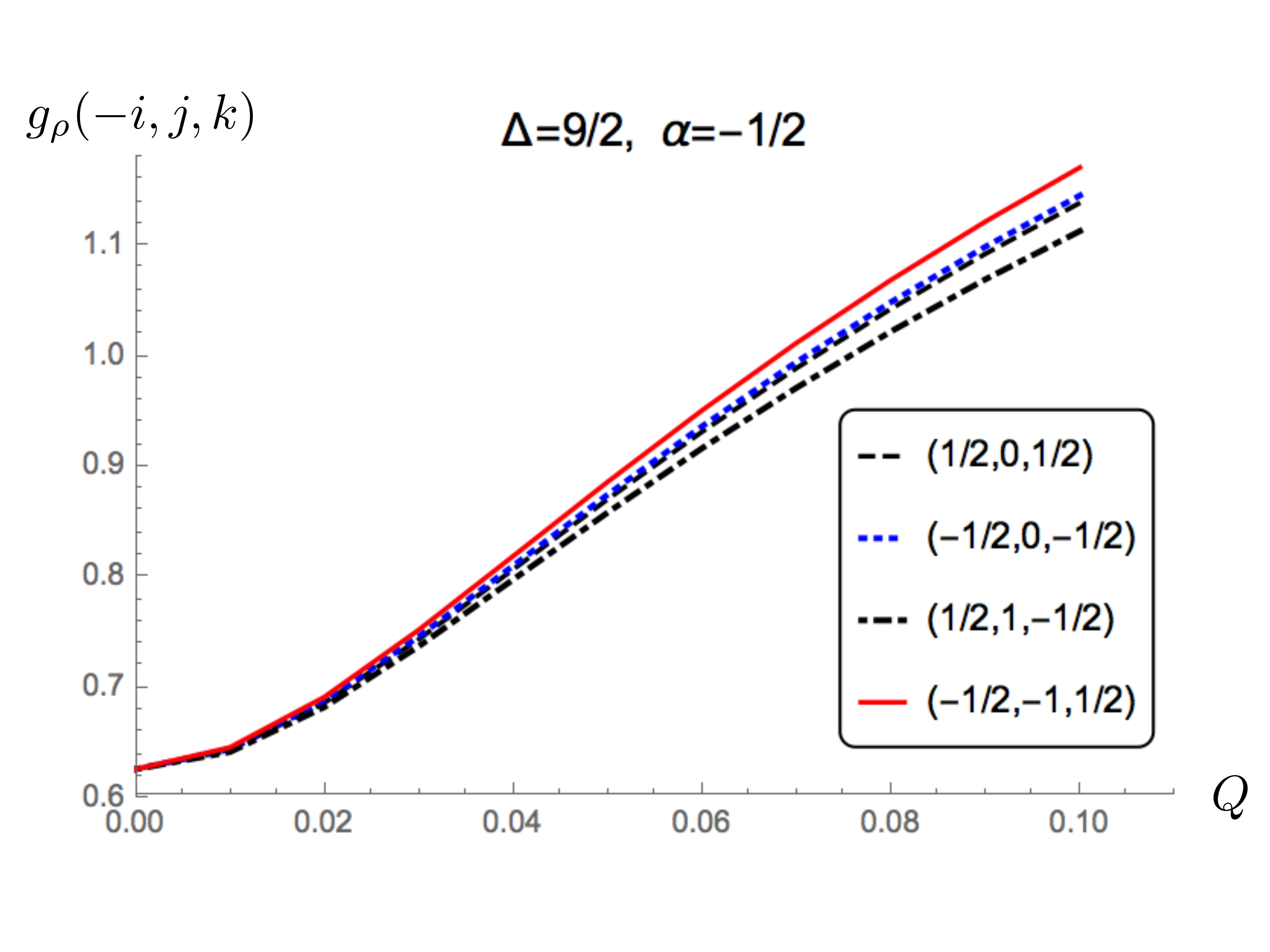}}
\hspace{0cm}
\subfigure[]
{ \includegraphics[angle=0,width=0.48\textwidth]{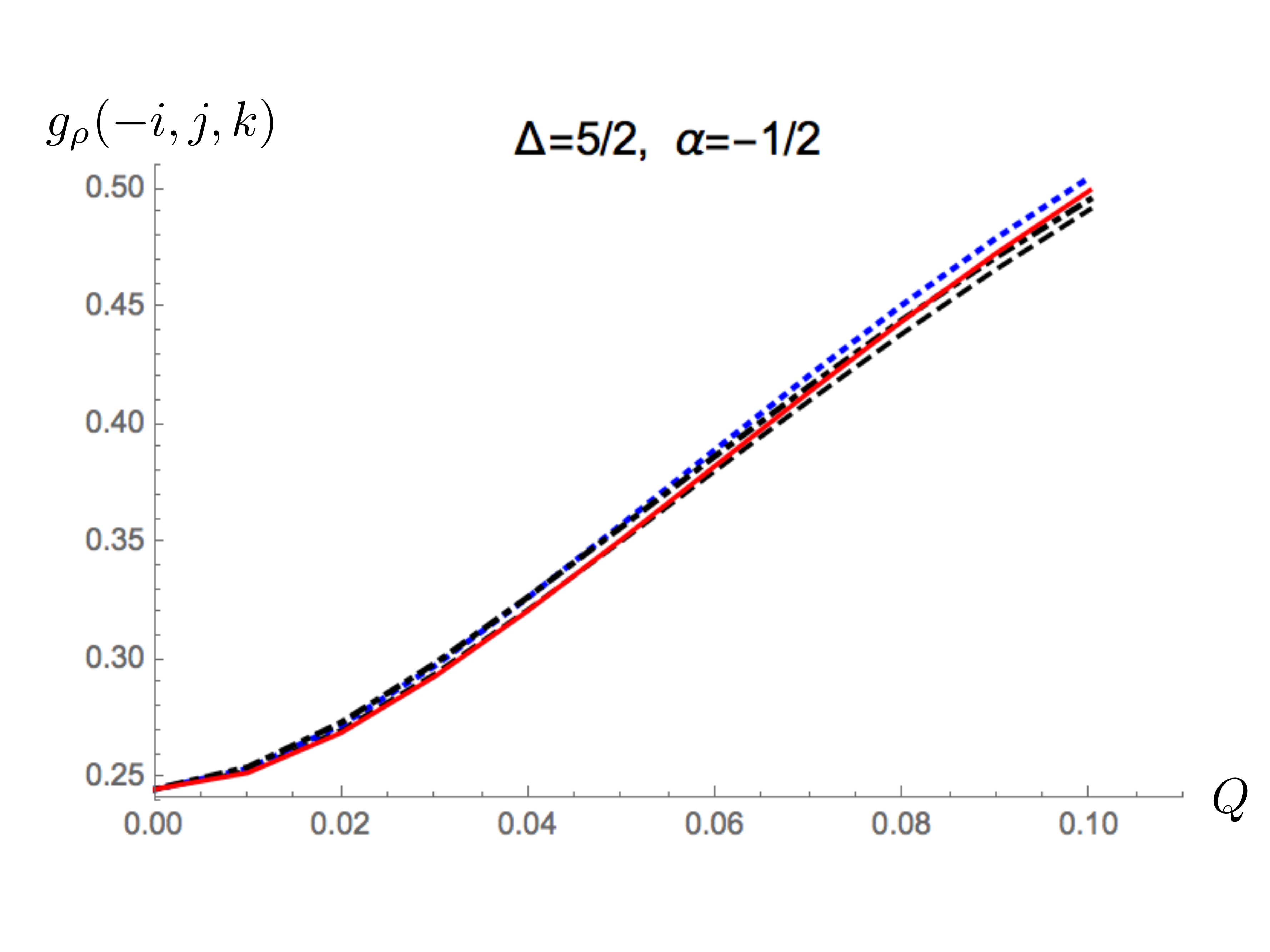}}
\vspace{-0cm}
\subfigure[]
{ \includegraphics[angle=0,width=0.48\textwidth]{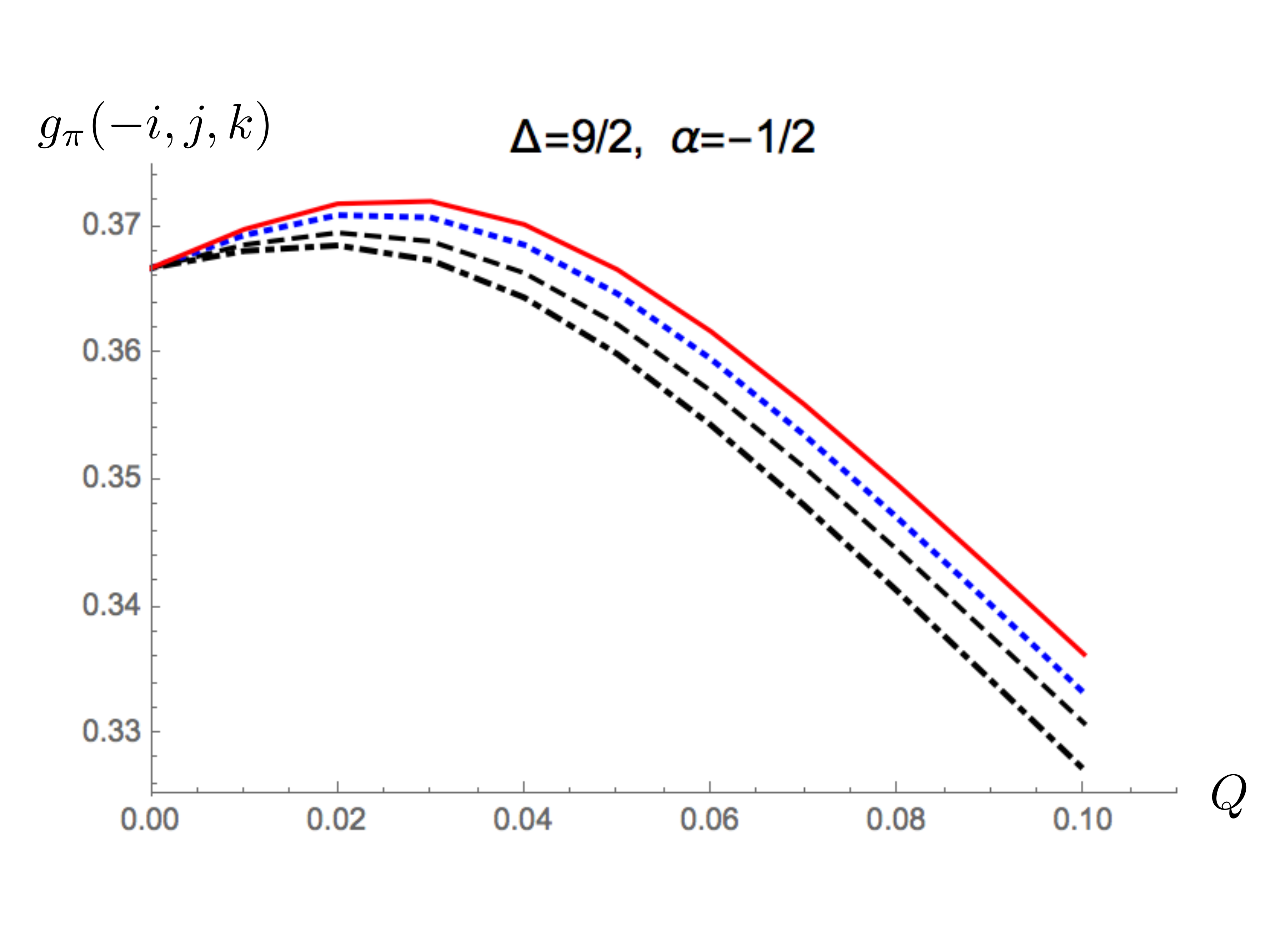}}
\vspace{-0cm}
\subfigure[]
{ \includegraphics[angle=0,width=0.48\textwidth]{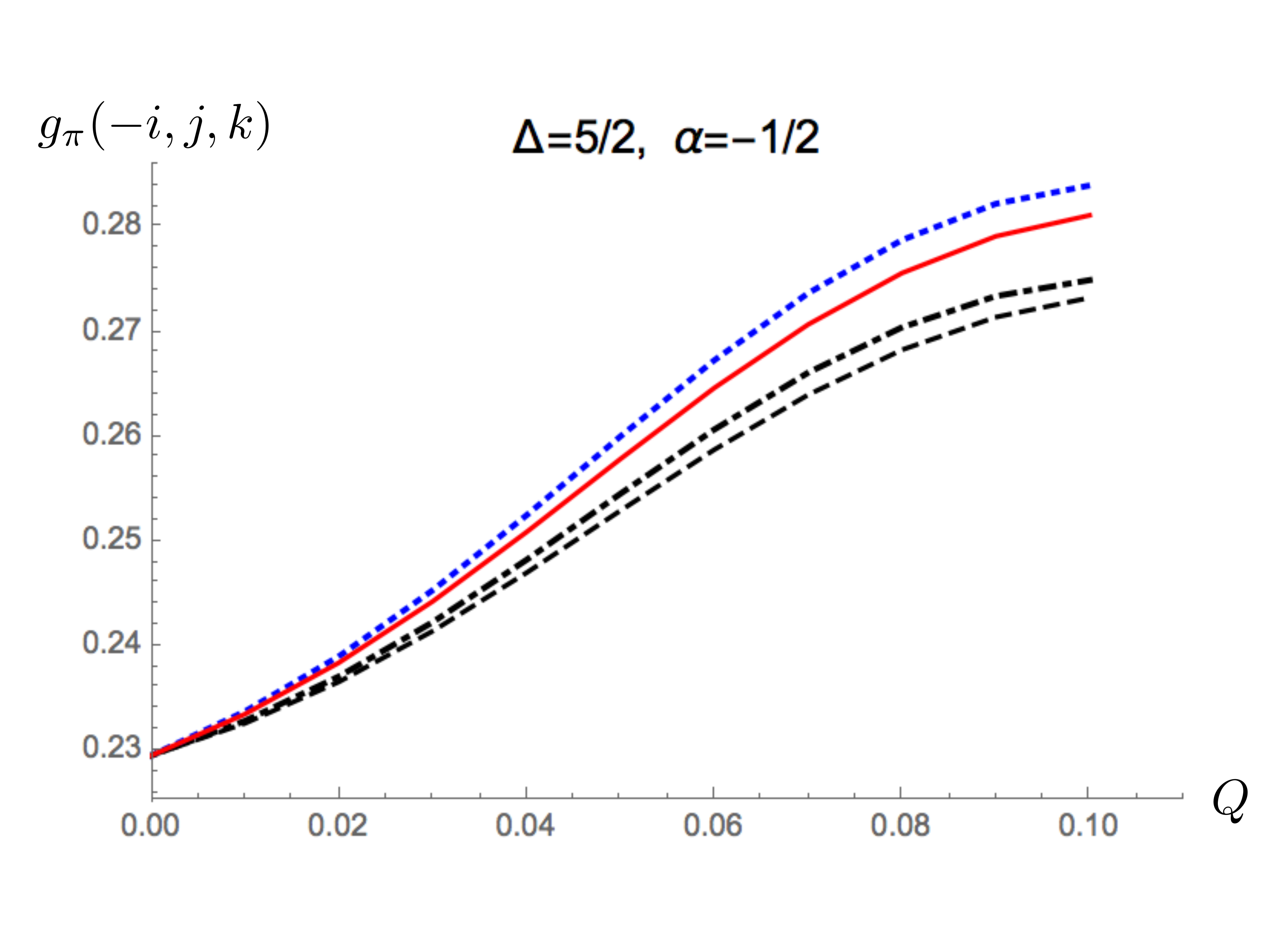}}%
\vspace{-0cm}
\subfigure[]
{ \includegraphics[angle=0,width=0.48\textwidth]{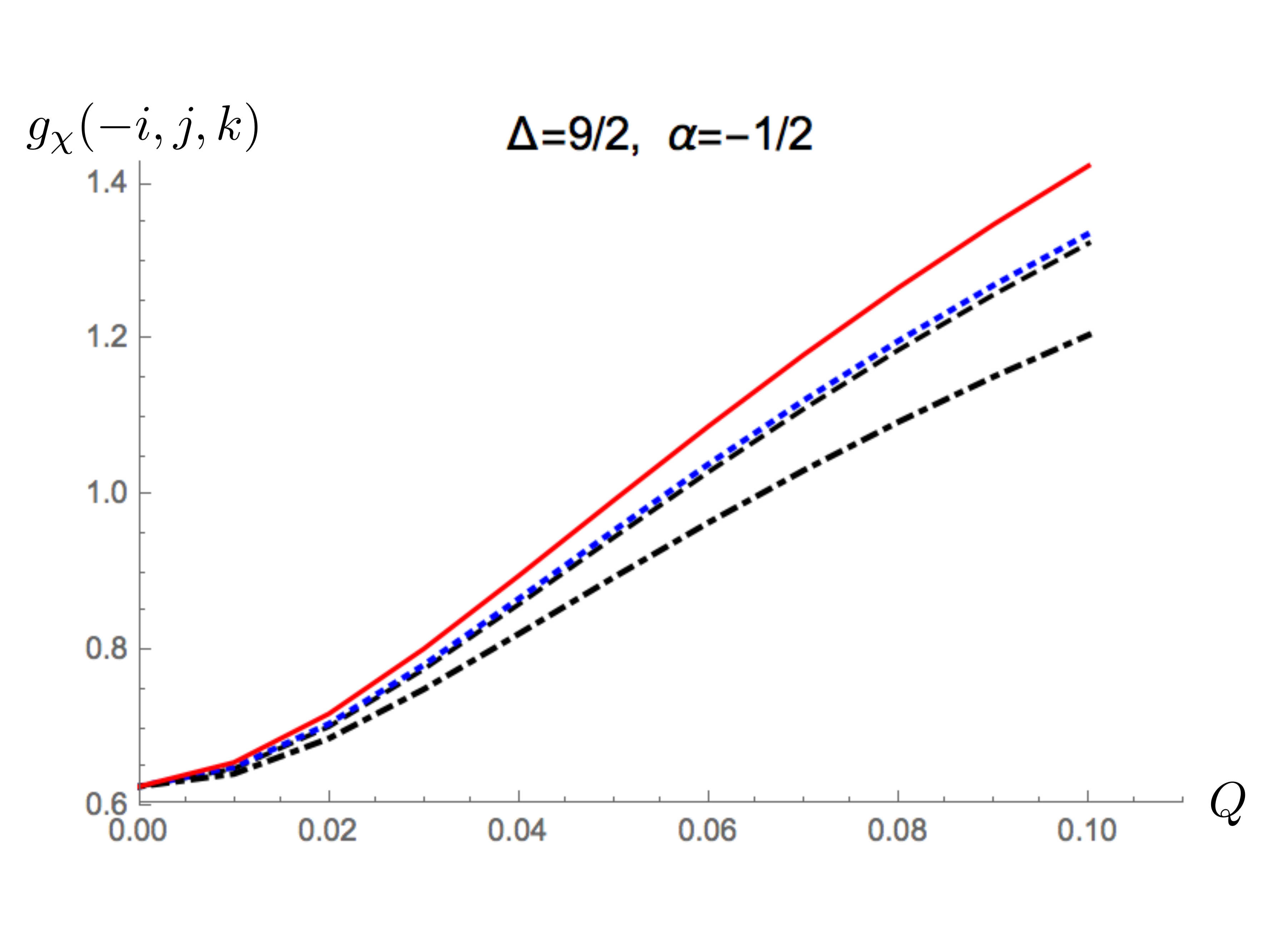}}
\vspace{-0cm}
\subfigure[]
{ \includegraphics[angle=0,width=0.48\textwidth]{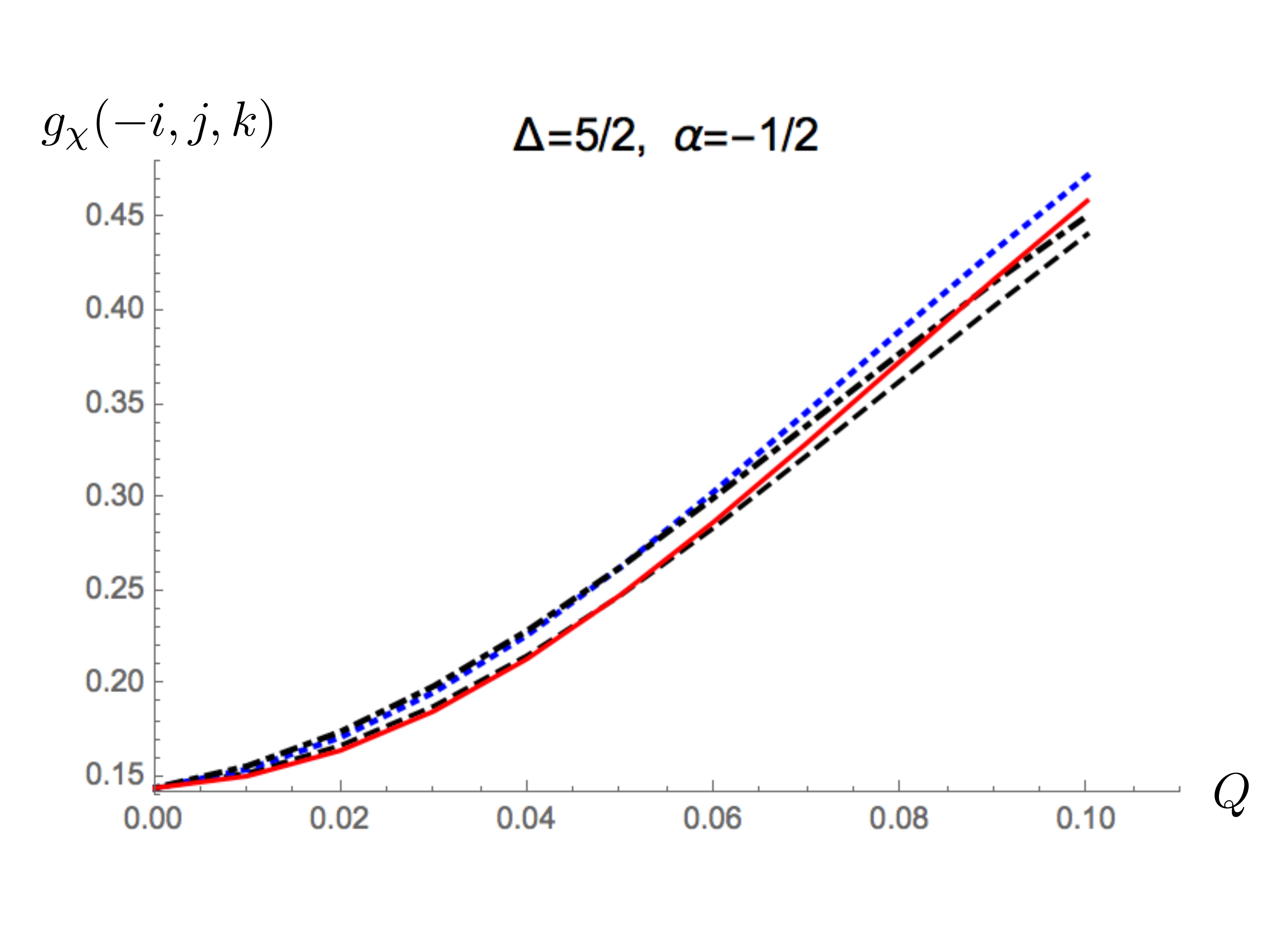}}
\vspace{-0cm}
\caption{\small  For $\alpha=-1/2$ the nucleon form factors with $\Delta=9/2$ ((a), (c), (e)) and $\Delta=5/2$ ((b), (d), (f)).}
\label{number}
\end{center}
\end{figure}



\section{Discussion}

By using the holographic technique, we investigated the nucleon's form factors relying on the density of the background nuclear medium. In order to imitate the nuclear medium holographically, we considered the thermal charged AdS space. Its dual field theory represents a nuclear medium classified by the quark number density and isospin charge. On this background geometry, we turned on the bosonic and fermionic fluctuations corresponds to mesons and nucleons, respectively. By solving linearized equations of motion for fluctuations, we found the mass spectrum of mesons and nucleons. And then, we derived their form factors by rewriting the bulk Yukawa coupling in terms of boundary fields corresponding to nucleons and mesons. 

For $\a=0$ where the numbers of proton and neutron are the same, we showed that the form factor is degenerate because the effect of the isospin interaction vanishes. For $\a \ne 0$, on the other hand, the asymmetry of the nucleon density leads to the nontrivial isospin interaction
and splits the degenerated form factor into four different cases relying on the isospin charges of nucleons and mesons. In all cases, the nucleon form factors describing the interaction with $\r$-meson and the $\chi$ field increase as the density of nuclear medium increases. On the other hand, the nucleon form factor caused by the interaction with the $\pi$ field increases slightly and then rapidly decreases.

In this work, we introduced the anomalous dimension as a free parameter. Our numerical result showed that the anomalous dimension reduces the strength of the form factor in the low density region. Anyway, it should be noted that the anomalous dimension must be appropriately explained by the renormalization group procedure. In the holographic setup, it can be accomplished by studying the holographic renormalization and its RG flow after regarding the gravitational backreaction of the bulk fluctuations. This is important to understand the baryonic physics, their spectra and interactions, in the nuclear medium. In addition, it would be important to know the structure of a neutron star. We leave this issue as a future work and hope to report more results in future works.

\vspace{1cm}

{\bf Acknowledgement}

This work was supported by the Korea Ministry of Education, Science and Technology, Gyeongsangbuk-Do and Pohang City. CP was also supported by Basic Science Research Program through the National Research Foundation of Korea funded by the Ministry of Education (NRF-2016R1D1A1B03932371). JHL was also supported by Basic Science Research Program through the National Research Foundation of Korea (NRF) funded by the Ministry of Education (NRF-2016R1A6A3A01010320).

\vspace{1cm}


\bibliographystyle{utphys}
\bibliography{ref1}{}

\end{document}